\documentclass[twocolumn,showpacs,amsmath,prd]{revtex4}
\usepackage[dvips]{epsfig}

\begin{document}

\title{Particle Acceleration through Multiple Conversions from 
Charged into Neutral State and Back}

\author{E.V. Derishev$^{1,2}$, F.A. Aharonian$^1$, V.V.
Kocharovsky$^{2,3}$, Vl.V. Kocharovsky$^2$}

\affiliation{
$^1$MPI f\"{u}r Kernphysik, Saupfercheckweg 1, D-69117 Heidelberg, Germany\\
$^2$Institute of Applied Physics RAS,
46 Ulyanov st., 603950 Nizhny Novgorod, Russia\\
$^3$Dept. of Physics, Texas A\&M University, College Station,
TX 77843-4242}

\begin{abstract}

We propose a new means for quick and efficient 
acceleration of protons and/or electrons in relativistic bulk flows. 
The maximum attainable particle energies are limited either by 
radiative losses or by the condition of confinement in the magnetic 
field. The new mechanism takes advantage of conversion of particles 
from the charged state (protons or electrons/positrons) into neutral 
state (neutrons or photons) and back. In most cases, the conversion 
is photon-induced and requires presence of intense radiation fields, 
but under special circumstances the converter acceleration mechanism 
may operate via other charge-changing reactions, for example, 
inelastic nucleon-nucleon collisions.
Like in the traditional,  ``stochastic'' (or diffusive) 
acceleration models, the acceleration cycle in the proposed scenario 
consists of escape of particles from the relativistic flow followed by their 
return back after deflection from the ambient magnetic field. The 
difference is that the charge-changing reactions, which occur during 
the cycle, allow accelerated particles to increase their energies in 
each cycle by a factor much larger than 2 and usually roughly equal 
to the bulk Lorentz factor squared. The emerging spectra of 
accelerated particles can be very hard and their maximum energy in 
some cases is larger than in the standard mechanism. This 
significantly reduces the required energy budget of the sources of 
the highest-energy particles observed in cosmic rays.  
% The proposed 
acceleration mechanism has a distinctive feature -- it unavoidably 
creates neutral beams, consisting of photons, neutrinos or neutrons, 
whose beam pattern may be much broader than the inverse Lorentz 
factor of the relativistic flow. Also, the new mechanism may serve as 
an efficient means of transferring the energy of bulk motion to 
gamma-radiation and, if the accelerated particles are nucleons, 
inevitably produces high-energy neutrinos at relative efficiency 
approaching $\gtrsim 50 \%$.

\end{abstract}


\maketitle

\section{Introduction}

The cosmic rays (CRs) have been studied for a long time (see, e.g., 
\cite{CRrew} for a review). There are many models of their (likely 
non-uniform) sources, but explaining the origin of the highest-energy 
particles ($\gtrsim 10^{20}$~eV) observed in CRs is still a 
challenging problem \cite{prd}. In general, two scenarios have been 
proposed for the particle acceleration in astrophysical environments. 
One is the acceleration by electric field in geometries, where this 
field is not perpendicular to magnetic field-lines, for example, in 
the vicinity of magnetized rotating neutron stars \cite{NSacc}. Very 
hard particle spectra may emerge in this way, but, because of the 
curvature losses inherent to such geometries, the upper limit to 
proton energy appears to be below $10^{20}$~eV. 

Another class of scenarios assumes gradual, ``stochastic'' (or 
diffusive) acceleration of charged particles through multiple 
reflections from inhomogeneities of magnetic field in the 
environments where large velocity gradients are present (see 
\cite{fermi} for a review). According to the generally accepted view 
(see, e.g., \cite{fermi,flow1,flow2}) this mechanism works equally 
well for shocks and shear flows, and the emerging spectra of 
accelerated particles are such, that only a small fraction of the 
total energy budget is contained in the most energetic particles.  

Furthermore,  there is a serious obstacle in achieving high energies of the 
accelerated particles -- their diffusive escape.  In the 
environments, where bulk velocities are sub-relativistic ($v \ll c$), 
it takes many reflections to increase the energy of accelerated 
particle twofold. In this case, the mechanism 
can work only in the largest objects in the Universe, like galaxy 
clusters or radio lobes and knots in active galactic nuclei (AGNs), 
and only under optimistic assumption about the diffusion coefficient 
for particles in magnetic field. For example, Bohm diffusion leads to 
particle escape on the timescale $R_0^2/r_g c$, where $R_0$ is the 
accelerator's size and $r_g$ the particle's gyroradius, while the 
acceleration occurs on the timescale $\sim (c/v)^2 r_g/c$. The 
maximum attainable energy is defined by equating these timescales:  
$r_g = (v/c) R_0 \ll R_0$. It is of the order of $10^{19}$~eV 
in the most favorable cases.

Ultrarelativistic shocks and shear flows are more promising, since 
the parameter $v/c \rightarrow 1$. Moreover,
the energy gain per cycle, i.e., at each reflection from the shock 
or from the shear flow boundary, can approach the 
factor of $\sim \Gamma^2$, where $\Gamma$ is the bulk Lorentz factor 
(measured in the upstream fluid frame in the case of shock). Such a 
scenario was suggested, e.g., as the dominant acceleration mechanism 
in gamma-ray bursts (GRBs) \cite{LevEich,Viet,Waxm}.  However, in the 
standard acceleration theory, the energy gain of $\sim \Gamma^2$ 
occurs only at the first cycle, while the subsequent ones result in 
energy gain of $\sim 2$ each \cite{BO,kirk}. 

The reason is that the shock catches 
up to the reflected particle (or the particle crosses the boundary of 
the shear flow) when its trajectory makes an angle of $\simeq 
2/\Gamma$ to the shock normal (or bulk velocity vector), while 
keeping the $\Gamma^2$ energy gain would require isotropisation. 

The smallness of this angle is apparent for the case 
of shear flow (see Fig.~1b). For shocks it is a consequence of 
relativistic motion (Fig.~1a). Indeed, the component of particle 
velocity along the shock normal becomes smaller than the velocity of 
the shock itself as soon as the angle between the particle's momentum 
and the shock normal grows larger than $1/\Gamma$. 
There is no apparent way to 
isotropize reflected particles unless a special structure in the 
magnetic field ahead of the shock is introduced, e.g., a 
counter-propagating shock.

\begin{figure}
\label{f1}
\scalebox{0.6}{\includegraphics{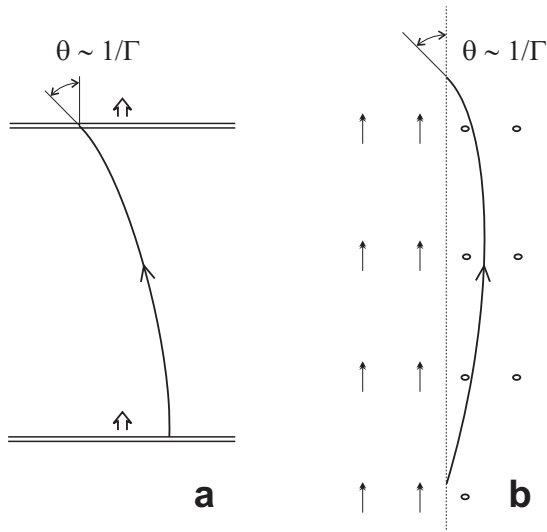}}
\caption{
\small The acceleration cycle in the standard mechanism for a shock 
(a) and for a shear flow (b). The thick solid line shows the 
particle's trajectory.  The magnetic field is perpendicular to the 
picture plane. The locations of the shock at the 
moments of particle escape from the shock and subsequent catch-up are 
shown as double lines. The shear flow boundary is shown by thin 
dotted line.} 
\end{figure}

However, a means to circumvent these limitations exists \cite{Rome} 
-- it can be done by switching particle's charge on and off at right 
times.  Paradoxically, interactions with photons, which have been 
always treated as dissipative  processs leading to degradation of 
particle energy, in fact play positive role: they allow 
(through the charge-changing particle conversion) to keep the 
$\Gamma^2$ energy gain up to the largest particle energies. There 
are also other types of conversion reactions, which we briefly 
discuss in the following section. One of the suggested below 
charge-changing schemes (electron-photon-electron reaction chain in 
GRB shocks) is independently considered in \cite{Stern}.

In order to outline the general picture, we intentionally skip some 
details, which are not essential for the proposed acceleration 
mechanism, but may change its quantitative characteristics. In 
particular, we assume that the magnetic field is either  
chaotic (turbulent) or uniform with field lines perpendicular to both 
the momentum of accelerated particle and the velocity of the flow,
and treat shocks and shear flows as one-dimensional discontinuities. 

We use the convention $F_{\nu} \propto \nu^q$ to define the spectral 
index $q$, where $F_{\nu}$ is the energy flux per unit frequency 
interval.

\section{The converter acceleration mechanism}

Two basic types of photon-induced conversion involve nucleons or 
electrons/positrons. Both cycles consist of two reactions:  
\begin{equation}
\label{pcycle}
p+\gamma \rightarrow n + \pi^{+} \qquad \mbox{and} \qquad 
n + \gamma \rightarrow p + \pi^{-}
\end{equation}
for proton cycle, 
\begin{equation}
\label{ecycle}
e^{\pm}+\gamma \rightarrow e^{\pm} + \gamma^{\prime} \qquad 
\mbox{and} \qquad 
\gamma^{\prime} + \gamma \rightarrow e^{+} + e^{-} 
\end{equation}
for electron cycle.
Here $p$, $n$, $\pi^{\pm}$, and $e^{\pm}$ denote proton, neutron, 
charged pions (positive and negative), positron and electron, 
respectively; $\gamma$ is a low-energy background photon and 
$\gamma^{\prime}$ the high-energy comptonized photon.

The second of reactions from the electron cycle (\ref{ecycle}) has a 
kinematic threshold $\Delta_e = 2\, m_e c^2$ in the 
center-of-momentum frame, where $m_e$ is the electron mass. 
Effectively, the first reaction also has 
the threshold $\simeq \Delta_e$, since at lower energies of incident 
photons the efficiency of energy transfer to the comptonized photon 
becomes much less than unity.  The reactions proceed differently 
depending on the background photon spectrum.  Soft spectrum blocks 
the electron cycle as the fraction of energy transferred to a 
comptonized photon is too small and there are few target photons
sufficiently energetic for the second reaction of the electron cycle. 
An example of soft spectrum could be a power-law with spectral index 
$q < -1$ or a narrow-band spectrum, like black-body or line emission, 
with typical photon energy $\bar{\varepsilon} \ll m_e^2 c^4/ 
\varepsilon_e$, where $\varepsilon_e$ is the electron energy. For
intermediate spectra (e.g., power-laws with indices $-1 < q < 
1$ or narrow-band spectrum with $\bar{\varepsilon} \sim m_e^2 c^4/ 
\varepsilon_e$), the comptonized photon takes about 1/2 of the 
electron (positron) energy, and in the consequent pair production 
event this energy is divided nearly in equal parts between the 
daughter electron and positron. The cross-section in both processes 
is $\sigma_{e,\gamma} \sim 10^{-25}$~cm$^2$.
For hard spectra ($q>1$ or $\bar{\varepsilon} \gg m_e^2 c^4/ 
\varepsilon_e$), these reactions proceed in the deep 
Klein-Nishina regime, i.e., the comptonization and pair-production 
cross-sections decrease inversely proportional to the square of the 
center-of-momentum energy (their ratio is 1:2) and almost all the 
energy of interacting particles is transferred to one of the daughter 
particles. In effect, the energy losses for the combined 
electron/photon particle become very gradual. This case is the 
closest to the pure conversion (the probability of charge change $p_c 
=1$) without accompanying energy losses provided the synchrotron 
emission is negligible.

The reactions from the proton cycle (\ref{pcycle}) (see, e.g., 
\cite{RM}) have kinematic threshold of $(m_{\pi} c^2 + m_{\pi}^2 c^2 
/2\, m_N) \simeq 150$~MeV in the nucleon rest frame ($m_{\pi} \simeq 
140$~MeV$/c^2$ is the charged pion mass and $m_N \simeq 
940$~MeV$/c^2$ the nucleon mass).  Side by side with the reactions 
(\ref{pcycle}) proceed other photopionic reactions with formation of 
neutral pion, which preserve nucleon's charge. They have roughly the 
same cross-section and should be considered as a background energy 
losses. The total photopionic cross-section rapidly increases with 
energy of incident photon and reaches maximum of $\sigma_{\pi} \simeq 
6\times 10^{-28}$~cm$^2$ at $\Delta_p \simeq 340$~MeV, which 
corresponds to formation of $\Delta$-resonance and should be 
considered as effective threshold.  Well above the resonance energy, 
the cross-section decreases and levels off at $\simeq 
10^{-28}$~cm$^2$.  The probability of charge change in a photopionic 
reaction is $p_c \simeq 1/3$ at the resonance and $p_c \simeq 1/2$ at 
the plateau. The inelasticity is $\simeq 0.2$ and about 0.5, 
respectively.

A competing photon-induced reaction is the process of creation of an 
electron-positron pair by a photon interacting with the electric 
field of a proton $p+\gamma \rightarrow p+ e^{-} + e^{+}$, which has 
the cross-section $\simeq 5\times 10^{-27}$~cm$^{2}$ and inelasticity 
$\simeq 10^{-3}$. With the decrease of spectral index $q$ of 
target photon field, this process becomes an increasingly important 
energy loss channel, and at $q \simeq -1.55$ the difference with the 
photopionic processes in the inelasticity and cross-section is 
exactly balanced by larger number of target photons (thanks to lower 
threshold). Anyway, the $p+\gamma \rightarrow p+ e^{-} + e^{+}$ 
process can be neglected for spectra with $q \gtrsim - 1.5$.

In dense environments and at relatively low nucleon energies, i.e., 
in the case where there are few target photons, the proton cycle 
proceeds through inelastic nucleon-nucleon collisions, for example,
\begin{equation}
\label{pcycle1}
p+p \rightarrow n + p + \pi^{+} \qquad \mbox{and} \qquad 
n + p \rightarrow p + p + \pi^{-}\, .
\end{equation}
The kinematic threshold for these reactions is $m_{\pi} c^2$ in the 
center-of-momentum frame, and the cross-section at energies $\gg 
m_{\pi} c^2$ is $\simeq 3\times 10^{-26}$~cm$^2$. The acceleration 
via inelastic nucleon-nucleon collisions could be important in 
GRB internal shocks, where the required column density of 
$\sim 10$~g/cm$^2$ is achieved \cite{insh}.

In both proton and electron cycles, one can consider an accelerated 
nucleon or electron/positron as a particle, which has both charged 
and neutral states. The acceleration cycle consists of three steps 
(see Fig.~2).  First, a charged particle in relativistic flow 
is converted into neutral state (point 1). Then, experiencing no 
influence from the magnetic field, it freely leaves the flow and 
propagates into ambient medium much further than if it were charged. 
Second, a transition from neutral to charged state occurs (point 2), 
which may be the spontaneous neutron decay. 

At this 
moment, the particles in the laboratory frame preserve beaming with 
the opening angle of $\sim 1/\Gamma$, which they had in the neutral 
state. The initial handicap allows particles to be deflected by an 
angle $\theta \gg 2/\Gamma$ before the encounter with the 
relativistic flow. The angular spreading of the 
particle beam in the laboratory frame means energy gain in the flow 
comoving frame, which is much larger than $2$ and amounts to 
$\Gamma^2$ in the case of full isotropisation. At the third step, the 
particles return to the flow (point 3) and their isotropisation in 
the comoving frame translates into the resulting energy gain in the 
laboratory frame.

\begin{figure}
\label{f2}
\scalebox{0.6}{\includegraphics{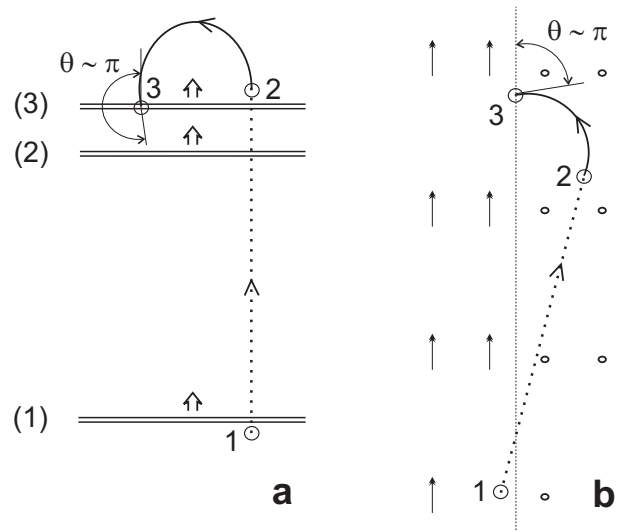}}
\caption{
\small The acceleration cycle in the converter mechanism for a shock 
(a) and for a shear flow (b). The particle's trajectory is shown by 
thick dotted line (neutral state) and thick solid line (charged 
state). The magnetic field is perpendicular to the picture plane.  
Numbered are the moments of particle conversion into neutral state, 
transition from neutral to charged state, and subsequent return to 
the flow. The locations of the shock at the corresponding moments 
are shown by double lines. The shear flow boundary is shown by thin 
dotted line.}  
\end{figure}

The main parameter, characterizing the efficiency of the converter 
mechanism, is the optical depth for interactions of accelerated 
particles (protons/neutrons or electrons/photons) $\tau = \sigma n 
D$, where $n$ is the number density of target particles (photons or 
nucleons) and $D$ the accelerator's size, both measured in the 
comoving frame, $\sigma$ the relevant cross-section. 

The optical depth is geometry-dependent.
In the case of a continuous outflow or a shock, produced by a central 
engine and subtending an angle $> 1/\Gamma$, one has for 
photon-induced reactions 
\begin{equation} 
\tau \simeq 
\frac{\sigma L(\varepsilon_*) \Theta^2}
{4\,\pi R c \varepsilon_*}.
\end{equation}
Here $L$ is the apparent luminosity per logarithmic frequency interval at 
photon energy $\varepsilon_* =2\, mc^2 \Delta/(\varepsilon \Theta^2)$,
where the interaction with target photons is the most efficient, 
$\varepsilon$ the energy of accelerated particle, $R$ the distance 
from the central engine, $\Delta$ and $m$ are the threshold and the 
mass of the particle for one of the possible cycles. 

The beaming 
angle of target radiation field is $\Theta \sim 1/\Gamma$ in the case 
where it is produced within the jet or by the shocked gas, and 
$\Theta \sim 1$ for the emission from broad-line regions in AGNs and 
the radiation scattered in the interstellar medium around GRBs. 
Intermediate cases, where $1/\Gamma < \Theta < 1$, are also possible. 
They include, for example, the radiation from inner parts of 
accretion discs in AGNs or the emission which accompanies the 
acceleration process (as explained in Sect.~V).
If the source of target radiation is transient, like a GRB, with 
duration less than $R \Theta^2/ c$, then its luminosity should be 
replaced by $E c/(R \Theta^2)$, where $E = \int L\, dt$. 

The most favorable for the converter mechanism conditions exist in 
AGNs and GRBs, where both 
ultrarelativistic flows and intense radiation fields are present 
(see \cite{AGNrew} and \cite{GRBrew} for reviews).  Other objects 
with relativistic outflows, e.g., stellar-mass microquasars, may also 
be able to accelerate particles via the converter mechanism. The only 
two prerequisites are sufficiently high conversion probability 
(which we specify below) and bulk Lorentz factor $\Gamma \gg 2$ in 
order to compensate the energy losses caused by conversion. 

Because of relatively large cross-section, the optical depth for 
photon-induced reactions is not a bottleneck for the electron cycle, 
but it could be a limiting factor for the proton cycle. Let us 
estimate the optical depth for photopionic reactions in three cases.
For AGN inner jets we obtain (taking into account only comoving 
photon fields with $\Theta = 1/\Gamma$)
\begin{equation} 
\label{AGNt}
\tau \simeq 10^{-1} 
\left( \frac{L (\varepsilon_*)}{10^{45}\, \mbox{erg/s}}\right)
\left( \frac{\varepsilon}{10^{18}\, \mbox{eV}}\right)
\left( \frac{10}{\Gamma}\right)^4
\left( \frac{10^{15}\, \mbox{cm}}{R}\right),
\end{equation}
where the apparent luminosity per logarithmic frequency interval $L 
(\varepsilon_*)$ depends on $\varepsilon$. The radiation from AGN
broad-line regions creates the optical depth 
\begin{equation} 
\label{AGNt-br}
\tau \sim 5\times 10^{-2} 
\left( \frac{L}{10^{44}\, \mbox{erg/s}}\right)
\left( \frac{10\, \mbox{eV}}{\bar{\varepsilon}} \right)
\left( \frac{10^{17}\, \mbox{cm}}{R}\right),
\end{equation}
which does not depend on the particle's energy: for all particles 
with energy $\varepsilon > 2\, mc^2 \Delta/ \bar{\varepsilon} 
\simeq 5\times 10^{16}$~eV the number of target photons is 
essentially constant because of their relatively narrow spectral 
distribution.  In GRBs, the optical depth due to comoving photons is 

\begin{equation} 
\label{GRBt}
\tau \simeq 3\times 10^{-3} 
\left( \frac{E (\varepsilon_*)}{10^{52}\, \mbox{erg}}\right)
\left( \frac{\varepsilon}{10^{16}\, \mbox{eV}}\right)
\left( \frac{100}{\Gamma}\right)^2
\left( \frac{10^{16}\, \mbox{cm}}{R}\right)^2.
\end{equation}
So, the conversion probability 
$p_{\rm cn} = [1-\exp(-p_c \tau)]$ in the proton cycle is 
usually, though not always, much smaller than unity. 

It should be 
noted, that effectively the probability of conversion of a neutron 
into a proton always exceeds
\begin{equation} 
\label{pmin}
p_{\rm cn}^{\rm (min)} = \frac{R m_N c}{t_n \varepsilon}
\simeq 3\times 10^{-2}
\left( \frac{10^{15}\, \mbox{eV}}{\varepsilon}\right)
\left( \frac{R}{10^{18}\, \mbox{cm}}\right)
\end{equation}
because of the neutron decay. Here $t_n \simeq 900$~s is the lifetime 
of free neutron. The spontaneous decay of free neutrons is important 
at small energies, especially during the first acceleration cycle, 
while at large energies the photon-induced conversion is more 
efficient.

The expressions (\ref{AGNt}), (\ref{AGNt-br}) and (\ref{GRBt}) cover 
all physically different situations. For example, one may use 
Eq.~(\ref{AGNt-br}) to estimate the optical depth in microquasars, 
where the target photons are produced by a hot corona having the size 
$R \sim 10^8$~cm and the luminosity $L \sim 10^{37}$~erg/s. The 
result is $\tau \sim 0.1$ at target-photon energy of $\sim 1$~keV, so 
that the proton cycle may operate in microquasars at energies 
$\gtrsim 3 \times 10^{14}$~eV. It is possible, therefore, that some 
contribution to the galactic CRs around the knee comes from 
microquasars. In any case, we see no problem in realization of the 
electron cycle in such objects.

\section{Energy gain}

Assuming that momenta of particles are isotropized in the comoving 
frame upon their encounter with the relativistic flow, one gets the 
average energy gain per cycle
\begin{equation}
g \simeq \frac{ \left( \Gamma \theta \right)^2}{2},
\end{equation}
where $\theta$ is the angle between the particle's momentum and the 
flow's velocity (deflection angle) at the moment of encounter. If 
the deflection angle is small, $\theta \ll 1$, then it grows linearly 
with distance $\ell$ (travelled by the particle in the charged state)
\begin{equation}
\theta = \theta_0 + \ell/r_g 
\end{equation}
in uniform magnetic field, while in chaotic field it behaves as
\begin{equation}
\label{defl-ch}
\left< \theta \right>  = \sqrt{\theta_0^2 + \ell \ell_c/r_g^2} 
\end{equation}
on average. Here $\theta_0$ is the initial deflection angle, measured 
at the time of conversion from neutral to charged state, $\ell_c < 
\ell$ the turbulence scale of the magnetic field, and $r_g$ the 
gyroradius of the particle, calculated as if the field were uniform.
Note, that the turbulence scale is implicitly defined by Eq. 
(\ref{defl-ch}) and, therefore, may depend on the particle's 
energy.

In the case of acceleration at the shock front, the shock catches up to 
the particle when the displacement of the particle along the shock 
normal, after it crossed the shock, becomes equal to the distance 
traveled by the shock front, i.e.,
\begin{equation}
\frac{\sqrt{\Gamma^2 -1}}{\Gamma}\,  (\ell_0 + \ell) =
\ell_0 \cos \theta_0 + \int_{0}^{\ell}\cos\theta \, d\ell^{\prime} \, ,
\end{equation}
where $\ell_0$ is the distance travelled by the particle in the 
neutral state after it left the relativistic flow, and we neglected 
the difference between the particle's velocity and the velocity of 
light.  

In the uniform magnetic field, the deflection angle is
\begin{equation}
\label{eq1}
\theta \simeq \left\{ \frac{3\, \ell_0}{r_g \Gamma^2} 
\left( 1- \Gamma^2 \theta_0^2 \right) \right\}^{1/3}
\sim \left( \frac{3\, \ell_0}{r_g \Gamma^2} \right)^{1/3}, 
\end{equation}
provided  $r_g/\Gamma \ll \ell_0 \ll r_g \Gamma^2$. For smaller 
initial displacement, one arrives to the result of 
standard theory $\theta \simeq 1/\Gamma$, whereas for larger 
displacement the deflection angle is $\theta \sim 1$. In the 
rightmost part of Eq. (\ref{eq1}) we assume that $\left( 1- \Gamma^2 
\theta_0^2 \right) \sim 1$ to simplify the algebra. This corresponds 
to ignoring particles, which escape propagating (in the comoving 
frame) nearly parallel to the shock plane. They are neither numerous 
nor energetically important.

If one substitutes $\ell_0$ by $R$ (or by $\Gamma D$ in the case of 
a small blob with the size $D < R/\Gamma$, ejected by a central 
engine), then the applicability limits give two critical energies 
(both measured at the end of acceleration cycle -- point 3 in Fig.~2):
\begin{equation}
\label{lim1}
\varepsilon_1 = \Gamma e B R  \qquad \mbox{and} \qquad
\varepsilon_2 = e B R, 
\end{equation}
where $e$ is the charge of the accelerated particle and $B$ the 
magnetic field strength. The acceleration proceeds with the maximum 
energy gain of $\sim \Gamma^2$ up to the energy $\varepsilon_2$, 
whereas above the energy $\varepsilon_1$ there is no advantage over 
the standard mechanism in the energy gain. At the same time, 
$\varepsilon_1$ is equal to the maximum energy, achievable in the 
standard mechanism. 

In the chaotic magnetic field 
\begin{equation} 
\label{eq2} 
\left< \theta \right> \simeq 
\left\{ \frac{2\, \ell_c \ell_0}{r_g^2 \Gamma^2} 
\left( 1- \Gamma^2 \theta_0^2 \right) \right\}^{1/4}
\sim \left( \frac{2\, \ell_c \ell_0}{r_g^2 \Gamma^2} \right)^{1/4}
\end{equation} 
for $r_g^2/\Gamma^2 \ll \ell_c \ell_0 \ll \Gamma^2 r_g^2$, and the 
critical energies are equal,
\begin{equation}
\label{lim2}
\varepsilon_1 = \varepsilon_2 = \Gamma e B \sqrt{R \ell_c}\, . 
\end{equation}

In the case of shear flow, the catch-up condition reads 
\begin{equation} 
\ell_0 \sin \theta_0 = 
\int_{0}^{\ell}\sin\theta \, d\ell^{\prime} \, .  
\end{equation} 
The deflection angle is
\begin{equation}
\label{eq3}
\theta \simeq \left( 2\frac{\ell_0 \theta_0}{r_g} \right)^{1/2} 
\sim \left( \frac{2\, \ell_0}{r_g \Gamma} \right)^{1/2} 
\end{equation}
for uniform field ($r_g/\Gamma \ll \ell_0 \ll r_g \Gamma$) and
\begin{equation}
\label{eq4}
\left< \theta \right> \simeq \left( \frac{3}{2}\frac{\ell_0 \ell_c
\theta_0}{r_g^2} \right)^{1/3} \sim \left( \frac{3}{2}
\frac{\ell_0 \ell_c}{r_g^2 \Gamma} \right)^{1/3}
\end{equation}
for chaotic field ($r_g^2/\Gamma^2 \ll \ell_0 \ell_c \ll r_g^2 
\Gamma$). 

There is a subtlety in realization of the converter mechanism if 
acceleration takes place at a shear flow boundary. If the charged 
particle reappears at a distance greater than $r_g$ (uniform magnetic 
field) or $r_g^2/\ell_c$ (chaotic field) from the flow boundary, then 
it should drift or diffuse back. It takes a time of the order of
$t_d = R^2/(\Gamma r_g c)$,
in the least favourable case of quasi-uniform magnetic field with 
characteristic spatial scale of $\sim R$, or
$t_d = R^2 \ell_c/(\Gamma^2 r_g^2 c)$,
in chaotic magnetic field. In both cases we assumed $\ell_0 \sin 
\theta_0 = R/\Gamma$. The solution of diffusion problem with a sink 
(the shear flow) in the half-space shows, that all particles 
eventually return to the flow, provided it persists for sufficiently 
long time. In reality, some of them are lost because the shear flow 
does not occupy the entire half-space, but the losses are negligible
if the spatial extent of the shear flow boundary is much larger than 
$\ell_0 \sin \theta_0$. This condition, for example, is satisfied for 
a conical jet with the opening angle $\gg 1/\Gamma$.

The synchrotron losses during the time $t_d$ might be a more 
serious problem.  One can neglect them if the energy loss 
rate $\dot{\varepsilon} = 4/9\, (e^2/mc^2)^2 B^2 (\varepsilon/mc^2)^2 
c$ multiplied by $t_d$ is less than the energy of accelerated 
particle. In the case of quasi-uniform magnetic field, this condition 
turns out to be energy-independent and, after simple algebra, one 
gets
\begin{equation} 
\label{drift}
R \gg \frac{4}{9} \frac{1}{\Gamma^4} \frac{e^2}{mc^2} 
\left( \frac{\varepsilon_2}{mc^2} \right)^3,
\qquad \mbox{i.e.,} \qquad 
R \gg \frac{R_{\rm opt}}{\Gamma^2}.
\end{equation} 
Here $R_{\rm opt}$ is the optimal size of an electromagnetic 
accelerator \cite{prd}, which is defined to minimize the amount of 
energy contained in the electric and magnetic fields. Let us recall, 
however, that the optimal size is originally defined for the 
accelerator itself (in this case -- for the interior of the shear 
flow), while Eq. (\ref{drift}) has to do with the external magnetic 
field. 

In a turbulent magnetic field, the synchrotron losses are the most 
significant for the least energetic particles, i.e., those having 
$r_g \sim \ell_c$. The losses can be neglected if
\begin{equation} 
\label{diff} 
R \gg \frac{R_{\rm opt}}{\Gamma^3},
\end{equation} 
where the optimal size is the function of critical energy 
$\varepsilon_2$, which is defined as if the magnetic field were 
uniform.

In order to be in agreement with the observed diffuse gamma-ray 
background, a typical source of the highest-energy CRs must satisfy 
the condition $R \gtrsim 0.2\, R_{\rm opt}$ \cite{prd}, so that the 
synchrotron losses during diffusion do not lead to any further
resrictions. The aforementioned statistical arguments are not 
applicable to individual abberrant accelerators, for which Eq. 
(\ref{drift}) or Eq. (\ref{diff}) should be considered as an 
additional limit.

The critical energies in the case of acceleration by a shear flow are 
\begin{equation} 
\label{lim3}
\varepsilon_1 = \varepsilon_2 = \Gamma e B R 
\end{equation} 
for uniform magnetic field and 
\begin{equation} 
\label{lim4}
\varepsilon_1 = \Gamma e B \sqrt{R \ell_c} \qquad \mbox{and} \qquad
\varepsilon_2 = \Gamma^{\frac{3}{2}} e B \sqrt{R \ell_c}
\end{equation}
for chaotic field. It is interesting to 
note, that in the latter case $\varepsilon_2 > \varepsilon_1$, i.e., 
the maximum energy achievable in the converter mechanism appears to 
be $\Gamma^{1/2}$ times larger than that in the standard mechanism. 
However, a particle can only attain this energy if it enters 
acceleration cycle with the energy $\varepsilon \sim 
\varepsilon_2/\Gamma^2$. A particle, entering the cycle with larger 
energy, will end up with smaller energy, in contrast to the other 
three cases.

Of course, every acceleration mechanism must obey the 
fundamental restrictions and constraints of classical electrodynamics 
\cite{prd}. Among them is the Hillas criterion \cite{Gr65,H84}
generalized for relativistic bulk flows, which is 
simply equivalent to the condition $\varepsilon < \varepsilon_1$, 
where $\varepsilon_1$ is the first critical energy from Eqs.  
(\ref{lim1}) and (\ref{lim3}). This is not a chance coincidence, but 
rather the consequence of the implied -- uniform -- magnetic field 
configuration. In a turbulent magnetic field, the maximum attainable 
energy is always lower than the fundamental limits, but the converter 
mechanism is less affected by the turbulence. Indeed, the 
magnetic field should be considered chaotic if its turbulence scale 
is less than the gyroradius of a particle at the beginning of the 
acceleration cycle. It means $\ell_c < r_g (\varepsilon_1)$ for the 
standard mechanism, but only $\ell_c < r_g (\varepsilon_2/\Gamma^2)$ 
for the converter mechanism. Thus, a particle accelerated in the 
standard way feels turbulence starting from larger scale, and a 
situation is possible, where the magnetic field should be treated as 
chaotic in the standard mechanism while being essentially uniform for 
the converter mechanism. The inequality $\varepsilon_2 > 
\varepsilon_1$ from Eq.~(\ref{lim4}) is, in fact, the consequence of 
slower decrease of critical energy with decreasing $\ell_c$ for the 
converter mechanism.

So far, the analysis was limited to the case of negligible 
synchrotron losses and $\tau \ll 1$.  The larger is the probability 
of conversion, the easier is acceleration in the converter mechanism.  
But if $\tau$ approaches unity, then the critical energies decrease: 
they are still given by Eqs. (\ref{lim1}), (\ref{lim2}), (\ref{lim3}), 
and (\ref{lim4}), but the value of $R$ should be replaced by the 
(energy-dependent) mean free path of the accelerated particle. 
Analogously, the synchrotron losses also decrease the critical 
energies; in this case $R$ should be replaced by the radiation 
length. 

In both cases the energy share of the accompanying radiation is 
non-negligible or even dominant, although the status of conversion 
losses is qualitatively different for the proton and electron cycles 
(see Sect.~V).

\section{Resulting particle distribution}

Let us consider a monoenergetic ($\varepsilon=\varepsilon_0$) beam of 
particles in the neutral state, escaping from the relativistic flow. 
If the probability of conversion per unit length, $\lambda$, is 
constant (i.e., we neglect the exponential decrease in the number of 
particles, assuming $\tau \ll 1$), then the distribution of charged 
particles over initial displacement $dN/d\ell_0 = \lambda$ translates 
into 
\begin{equation}
\frac{dN}{d \varepsilon} \propto 
\frac{dN}{d \ell_0} \frac{d\ell_0}{d g} \propto 
\lambda \left( \theta \frac{d\theta}{d\ell_0} \right)^{-1}\, ,
\end{equation}
where the particle's energy is simply proportional to the energy gain, 
$\varepsilon = g \varepsilon_0$.
It is a power-law distribution, but all the particles with 
displacement larger than the applicability limits of Eqs. (\ref{eq1}), 
(\ref{eq2}), (\ref{eq3}), and (\ref{eq4}) have the constant energy gain 
of $\sim \Gamma^2$, so that in general a delta-function is added to 
the resulting particle distribution at its high-energy end. In their 
power-law parts, the distributions emerging in uniform and chaotic 
field, respectively, are
\begin{equation}
\frac{dN}{d \varepsilon} \propto \varepsilon^{1/2}
\qquad \mbox{and} \qquad
\frac{dN}{d \varepsilon} \propto \varepsilon
\end{equation}
for acceleration at the shock front, and 
\begin{equation}
\frac{dN}{d \varepsilon} = {\rm const}
\qquad \mbox{and} \qquad
\frac{dN}{d \varepsilon} \propto \varepsilon^{1/2}
\end{equation}
for acceleration at the boundary of shear flow. These spectra are 
extremely hard. In practice, they can be considered delta-functions, 
so that the resulting distribution is defined mainly by the spectrum 
of injected particles.

\begin{figure}
\label{f3}
\scalebox{0.3}{\includegraphics{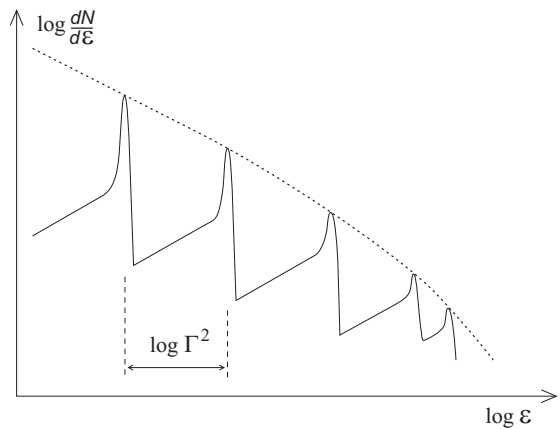}}
\caption{
\small The particle distribution (solid line), resulting from a 
monoenergetic injection. The dips in the distribution preserve 
themselves until the width of the injection spectrum is larger than 
${\rm log} \Gamma^2$ in logarithmic units.} \end{figure}

After a few cycles, a seesaw-shaped spectrum (Fig.~3) is formed from 
the initial monoenergetic distribution. The number of particles 
emerging from a cycle, $N^{\prime}$, is related to the number of 
particles entering it as $N^{\prime} = k N$, where 
\begin{equation}
k = p_{\rm esc} \prod_i p_{cn}^{(i)}
\end{equation}
is the overall probability of conversion, which normally occurs twice 
per cycle (points 1 and 2 in Fig.~2), multiplied by the probability 
of escape from the relativistic flow after the first conversion 
$p_{\rm esc}$, which is the relative number of particles moving 
upstream with respect to the shock front or the shear flow boundary.  
Assuming that the particle distribution is isotropic just before the 
conversion into neutral state, one obtains $p_{\rm esc} = 1/3$ for a 
strong relativistic shock, where the shocked fluid moves at the speed 
$c/3$ away from the shock plane, and $p_{\rm esc} = 1/2$ for a shear 
flow, where the fluid velocity is parallel to the boundary.  Within 
one acceleration cycle, the probabilities $p_{cn}^{(1)}$ and 
$p_{cn}^{(2)}$ are of the same order, except for the case where 
$p_{cn}^{(2)}$ is limited by Eq.~(\ref{pmin}).

There is a simple case, where $k$ remains constant in the course of 
acceleration. This happens in both electron and proton cycles if the 
spectrum of target photons is a power-law with spectral index $q=0$ 
or (for the proton cycle only) in the case of a narrow-band 
target-photon field. Then the envelope of the particle distribution 
shown in Fig.~3 is a power-law $dN/d \varepsilon \propto 
\varepsilon^{-\alpha}$, and the index $\alpha$ can be obtained in the 
following way.  

Since the particle energies are related as 
$\varepsilon^{\prime} = g \varepsilon$, we find that
\begin{equation}
\left( \frac{dN}{d \varepsilon} \right)^{\prime} = \frac{k}{g}\,
\frac{dN}{d \varepsilon},
\end{equation}
where primed is the distribution emerging from the cycle.
Taking logarithm of both parts, we obtain
\begin{equation}
-\alpha \ln g = \ln \left( \frac{k}{g} \right)
\qquad \Rightarrow \qquad \alpha = 1- \frac{\ln k}{\ln g}.
\end{equation}
If the probability of passing through the acceleration cycle $k$ 
is larger than the inverse energy gain $1/g$, then the spectral index 
is $\alpha < 2$, i.e., most of the energy content is at the 
high-energy end of the distribution. Both AGNs and GRBs can satisfy 
this requirement (see Eqs. \ref{AGNt}, \ref{AGNt-br} and \ref{GRBt}) 
and, therefore, can be efficient producers of the highest-energy 
cosmic rays. 

In the energy range, where the efficiency condition for the 
converter mechanism is satisfied, this mechanism is the dominant 
source of accelerated particles -- just because it provides a 
spectrum harder than the one resulting from the standard mechanism. 
At the same time, there is no actual threshold in the conversion 
probability:  the converter mechanism can function even at 
$p_{cn}\rightarrow 0$, but the number of the accelerated particles in 
this case is extremely depleted at high energies. Quantitative 
studies of the emerging CR spectrum must take into account the 
dependence of the conversion probability on the energy of accelerated 
particles. Hence, the precise solution can be obtained only in the 
self-consistent approach, which includes the 
effect of the accompanying emission, discussed in the following 
section.

\section{The accompanying emission}

Acceleration of particles via the converter mechanism is inevitably 
accompanied by gamma-ray and -- in the case of nucleon conversion 
-- by neutrino emission. The conversion losses in the electron cycle, 
which are the result of non-optimal inverse Compton scatterings, do 
not lead to irreversible energy drain. Instead, all the energy 
remains in photons or electrons (positrons) and, as they both 
can in principle participate in the acceleration cycle, may be used 
to inject new particles for acceleration. Thus, even in the case 
$\tau \gg 1$ the influence of the conversion losses is reduced to 
the decrease of maximum attainable energy and the increase in the 
level of synchrotron emission by rising the number of particles 
involved in the acceleration. 

On the contrary, the conversion losses in the proton cycle act as a 
true energy sink: the energy spent for the pion production  never 
comes back to the accelerated nucleons. Neutrinos, $e^{-}e^{+}$-pairs 
and gamma-rays, which are the decay products of pions, are copiously 
produced as byproducts of the proton acceleration cycle. Neutrinos 
carry away about one-half of the energy of accelerated nucleons and, 
because they freely escape from the acceleration site, their spectrum 
copies that of the nucleons, but scaled down in energies by about an 
order of magnitude.  The gamma-rays and pairs, which carry 
approximately the same energy as neutrinos, are reprocessed through 
the electromagnetic cascade:  the photons are absorbed in two-photon 
pair production process, electrons and positrons cool in the magnetic 
field, producing synchrotron radiation, i.e., another generation of 
photons, etc.

The hard spectrum of primary electrons and positrons means that they 
form a standard cooling distribution, $d N_e/ d \gamma_e \propto 
\gamma_e^{-2}$, and their synchrotron spectrum is a power-law with 
(photon) index $-3/2$. The synchrotron photons spawn another 
generation of pairs, which cool to form a distribution with index 
$-5/2$. The cascade comprises several steps like this, so that the 
photon spectrum eventually converges to the power-law with index 
$-2$.  However, at low energies the acceleration site is transparent 
for the photons, and therefore pairs are not injected below some 
energy, which leads to a break in the spectrum; below the break the 
photon spectrum preserves its original spectral index $-3/2$. The 
location of the break can be found in a self-consistent way: 
\begin{equation}
\varepsilon_{\rm br}  \simeq 0.5 \frac{\hbar e B}{\Gamma m_e c}
\left( \frac{\varepsilon_{\rm tr}}{2\, m_e c^2} \right)^2,
\end{equation}
where $\hbar$ is the Planck constant. The threshold photon energy 
$\varepsilon_{\rm tr}$ is defined so that
the optical depth for two-photon pair production,
\begin{equation}
\tau_{\rm \gamma \gamma} (\varepsilon_{\rm tr}) = 
\frac{\sigma_{\rm \gamma \gamma}}{\sigma_{\pi}} \tau_{\pi} 
\frac{\varepsilon_{\rm br} \varepsilon_{\rm tr}}{2\, \Gamma^2 m_e^2 
c^4} \left( \frac{\varepsilon_{*}}{\varepsilon_{\rm br}} 
\right)^{1/2},
\end{equation} 
is equal to unity, where $\sigma_{\rm \gamma \gamma}$ is the 
cross-section of two-photon pair production and $\tau_{\pi}$ the 
optical depth for photopionic reactions at photon energy 
$\varepsilon_{*}$. Straightforward calculations yield 
\begin{equation}
\label{Ebr}
\varepsilon_{\rm br}  \simeq 
0.5 \frac{\sigma_{\pi}}{\sigma_{\rm \gamma \gamma} \tau_{\pi}}
\sqrt{\frac{\hbar e B}{m_e c}\, 
\frac{\Gamma \varepsilon_N m_e^2 c^2}{\Delta_p m_N}}.
\end{equation}

Substituting $B = 10^{3}$~G, $\varepsilon_N = 10^{19}$~eV (the energy 
of accelerated nucleons), $\Gamma = 1000$, and $\tau_{\pi} = 
10^{-2}$, -- parameters, reasonable for GRBs, -- we obtain the break 
energy $\varepsilon_{br} \sim 100$~keV, which is similar to the 
really observed in GRBs spectral features. Within this picture, 
however, there is no simple way to explain the observed relative 
stability of $\varepsilon_{\rm br}$ whereas the parameters entering 
Eq.~(\ref{Ebr}), especially $\tau_{\pi}$, may vary by orders of 
magnitude. 

Unlike the conversion losses, the synchrotron emission is not tightly 
related to the acceleration process as such. It may have a negligible 
effect, especially for protons, but also may be the main energy loss 
channel. The detailed analysis of the properties of accompanying 
synchrotron radiation is beyond the scope of this paper, but we point 
out two distinctive features. 

First, the maximum energy of 
synchrotron photons for the converter mechanism is $\Gamma^2$ times 
larger than for the standard one. The existence of such an 
energy limit is easy to see for the standard mechanism  
(following the arguments of \cite{GFR}). 
The acceleration cycle in 
this case lasts $\sim r_g/c$ and the energy increment is $\sim 
\varepsilon$, which gives the acceleration rate $\dot{\varepsilon} 
\sim \varepsilon c/r_g$. The maximum admissible rate of synchrotron 
losses is just the same, so that the particle's energy is limited by 
the following inequality:
\begin{equation}
\label{syloss}
\frac{4}{9} \left( \frac{e^2}{mc^2} \right)^2 B^2
\left( \frac{\varepsilon}{mc^2} \right)^2 <
\frac{\varepsilon}{r_g}.
\end{equation}
Then, simple calculation yields the maximum energy of synchrotron 
photons (ignoring relativistic dipole radiation caused by small-scale 
inhomogeneities of magnetic field):
\begin{equation}
\varepsilon^{\rm (max)}_{\rm sy}  \sim 
0.5 \frac{\hbar e B}{mc} 
\left( \frac{\varepsilon}{mc^2} \right)^2 \sim
\frac{\hbar c}{e^2} mc^2 \simeq 137\, mc^2\, .
\end{equation}

Doppler boosting gives additional factor $\Gamma$, and in  a 
turbulent magnetic field with the spatial scale of the turbulence 
less than $\ell_c^{\rm cr} = m^2 c^4 /\sqrt{e^5 B^3}$ another factor 
$(\ell_c/\ell_c^{\rm cr})^{2/3}$ applies. 

The same reasoning is valid for the converter mechanism, in which the 
cycle duration is $\geq r_g/c$ and the energy increment is $\leq 
\Gamma^2 \varepsilon$. Consequently, the analog of Eq.~(\ref{syloss}) 
gives $\Gamma^2$ times larger limit for the energy of accelerated 
particle, which translates into factor $\Gamma^4$ in the expression 
for the energy of synchrotron photons. When the accelerated particle 
enters the relativistic flow being close to the limiting energy, the 
synchrotron emission is so efficient, that the particle loses almost 
all its energy before it is deflected by an angle $\sim 1/\Gamma$.  
Thus, the resulting synchrotron emission is beamed backwards in 
the flow comoving frame. In the laboratory frame it appears 
redshifted by the factor $\Gamma$, in contrast to the standard 
mechanism, in which the synchrotron emission is blueshifted in the 
laboratory frame by the same factor $\Gamma$. Thus, for an observer 
resting in the laboratory frame, the maximum energy of synchrotron 
photons accompanying the converter-acceleration is by the factor of 
$\sim \Gamma^2$ {\it larger} (recall the difference of $\sim 
\Gamma^4$ in the comoving frame), than in the standard mechanism.  
Moreover, this highest-energy synchrotron radiation is 
quasi-isotropic in the laboratory frame, which is another distinctive 
feature of the converter mechanism. The latter phenomenon has a 
nature similar to the effect of beam-pattern broadening for 
the inverse Compton radiation of electrons in front of relativistic 
shock \cite{rtrd}. Generally speaking, the converter mechanism makes 
neutral beams of all kinds (photon, neutrino and neutron beams) 
broader than $1/\Gamma$, so that they can be seen even if the jet 
that produced them is not pointing towards the observer.

The accompanying electromagnetic emission of any origin can itself 
provide photons for the conversion reactions, which may give rise to 
radiative instabilities analogous the instability driven by $p + 
\gamma \rightarrow p + e^{-} + e^{+}$ process, discussed in 
\cite{BH}.

\section{Discussion}

The key parameter of the converter mechanism -- the probability of 
conversion -- varies from one cycle to another or even within cycles. 
It increases in the course of acceleration (because a
more energetic accelerated particle interacts with less energetic and 
more abundant target particles) and with increase of the angle 
between the particle's trajectory and the velocity of the flow. The 
first effect is negligible for nucleon-nucleon collisions, while the 
second is partially or even completely compensated by smaller 
distance travelled at large angles. There is, however, a general 
trend of increase of the probability of conversion towards higher 
energies of accelerated particles; this may even block further 
acceleration if $p_{\rm cn}$ approaches unity. Nevertheless, with the 
expansion of the flow the density of target particles drops, leading 
to the decrease of the probability of conversion, and the 
acceleration resumes.

It means, that the converter mechanism is capable of self-tuning. The 
only thing which is required for this mechanism to be efficient is 
that the probability of passing through the acceleration cycle $k$ is 
larger than $1/\Gamma^2$ somewhere along the flow. It does not matter 
how large is the probability: in the case $k \rightarrow 1$ the 
particles are preaccelerated in the region of high optical thickness 
and then are further accelerated when (or where) the flow becomes 
optically thin for them. Of course, the question how much energy is 
wasted for conversion losses during this intermediate phase is open.

On the other hand, the condition $k > 1/\Gamma^2$ places some 
restrictions on the sources where the proton cycle is readily 
realizable starting from thermal protons. If a source is known to be 
bright at hard gamma-rays, one can conclude that the source's 
compactness, calculated for the corresponding target photons, is low, 
so that the probability of proton-to-neutron conversion is $k 
\lesssim \sigma_{\pi}/\sigma_{\gamma} \simeq 3\times 10^{-3}$. Thus, 
the source should have $\Gamma \gtrsim 20$ -- a condition, satisfied 
by GRBs and many AGNs. Moreover, this condition is not obligatory for 
the sources, where the converter mechanism starts from preaccelerated 
high-energy protons, which require only low-energy target photons.

We expect that the converter mechanism never operates alone and the 
standard Fermi-type mechanism competes with it. Because the 
particles spend some time in the neutral state, the converter 
mechanism has smaller acceleration rate at low energies. But, close 
to the limiting energy ($\varepsilon_1$), the durations of 
acceleration cycle in both mechanisms approach the same value $R/c$, 
so that the average acceleration rates are roughly equal. Except for 
the case of acceleration in a shear flow with chaotic ambient 
magnetic field, the mechanisms have just the same absolute, i.e., 
ignoring radiative losses, energy limit for the accelerated 
particles. However, in the converter mechanism, this energy is 
attained in much fewer steps, with potentially much more particles 
survived. Moreover, the particles -- when converted into neutral 
state -- can escape from regions located deep inside the relativistic 
flow, which further reduces irreversible particle losses in the 
downstream. So, at the highest energy part of the distribution almost 
all particles are produced by the converter mechanism, regardless of 
its performance at low energies.

Now let us consider the phase, when produced CRs leave the 
accelerator. Within the framework of the converter mechanism, the 
particles can escape in the form of neutral beam -- an easy way, 
which causes no problem and requires nothing but a sufficiently high 
conversion probability.  In the standard scheme, the escaping 
particles are charged and inevitably must form an expanding turbulent 
outflow. This causes considerable adiabatic losses, which can hardly 
be controlled.  Another advantage of the converter mechanism is its 
greater tolerance for non-uniform magnetic fields, as discussed in 
Sect.~III. Since the magnetic field turbulence has strictly negative 
effect on the maximum attainable energy, the converter acceleration 
mechanism may have a larger cut-off energy. 

Unlike the standard mechanism, the converter mechanism in many cases 
does not need any special particle injection. By definition of 
acceleration cycle, even the particles resting either in the 
laboratory or in the flow frame serve as injection as soon as they 
are converted into neutral state. Sometimes however, e.g., in AGNs, 
the probability of conversion is too small at low energies to provide 
sufficient injection. Then, the converter mechanism comes to depend 
on the standard mechanism, which produces preaccelerated particles 
(at energies $\gtrsim 10^{15}$~eV in the case of AGNs).

The photopionic processes in AGNs and GRBs leading to copious 
production of neutral particles have been extensively discussed in 
the literature (e.g., \cite{RM,AD}), in particular in the context of 
predictions of fluxes of high energy neutrinos from these objects.  
The main reasoning was that these objects are postulated to be 
efficient sources of high-energy CRs and they are surrounded by dense 
photon fields. Therefore, nucleon-photon interactions may 
lead to detectable neutrino fluxes. Our point of view is essentially 
different: if one wants to accelerate CRs in these objects to highest 
energies, then the converter mechanism is likely the most efficient 
way to do this, and the neutrino emission appears as a natural and 
unavoidable byproduct of this mechanism.

Further questions, absent in the test particle problem, arise in the 
self-consistent approach. In particular, the converter mechanism 
strongly alters hydrodynamics of relativistic shocks or shear flows. 
We have shown that they can accelerate nucleons with the efficiency 
of energy transfer approaching 100\%. At such a high efficiency, 
almost all the available energy is transferred to the end of the 
particle distribution, and hence the ultra-energetic particles 
contribute most of the inertia of the relativistic flow (shocked 
gas). Under such circumstances, the idea of a shock as a 
discontinuity becomes meaningless, since the gyroradius of 
dynamically most important particles is comparable to the size of the 
system.  

This limits applicability of the test particle approach, since it is 
only valid for the particles whose gyroradius is larger than the 
width of the shock or the shear flow boundary. The similar problem 
exists also in the standard mechanism, which is, moreover, more 
sensitive to it. Indeed, in the converter mechanism an accelerated 
particle may cross the shock or the shear flow boundary while being 
in the neutral state, hence relaxing the requirements to their 
sharpness. 

The absence of a true shock is an obstacle for the electron cycle, 
which would otherwise be a very efficient way of direct acceleration 
of electrons. When the mean free paths of photons and pairs are 
small compared to the spatial scale of velocity gradient, the direct 
acceleration of electrons probably cannot rival the efficiency of 
production of secondary pairs via nucleon acceleration. One of the 
implications is that at least in some, particularly bright, GRBs the 
main source of the observed gamma-ray radiation is the pion-induced 
cascade.

Energetic neutrons (or photons) escape from the relativistic flow and 
their decay or interaction products can disturb the ambient medium to 
the extent that it starts to move with an ultrarelativistic speed. An 
apparent consequence of this is the distortion of beam pattern of the 
source, while the influence on the limiting energy of the accelerated 
particles needs further investigation. On one hand, the decrease in 
the Lorentz factor contrast lowers the limiting energy. On the other 
hand, enhancement of the ambient magnetic field, either via 
compression in the shock or by means of various instabilities 
triggered by neutron decay products in the ambient plasma, has the 
opposite effect.

\section{Summary}

In this paper we suggest and analyze a new acceleration mechanism, 
which operates via continuous conversion of accelerated particles 
from charged into neutral state and back.
The proposed converter mechanism is efficient for acceleration of 
both protons and electrons (positrons). It is capable of producing 
the highest-energy ($\gtrsim 10^{20}$~eV) cosmic rays in either GRB 
or AGN environments. With a much lower energy limit, the mechanism 
can possibly operate in microquasars as well. In the regions of high 
optical thickness, the converter mechanism is an efficient means of 
transferring the kinetic energy of bulk relativistic flow to the 
accompanying radiation, which could explain, for example, the origin 
of GRB emission. Some peculiarities of the accompanying emission can 
be a tell-tale sign of the converter mechanism. For example, the 
production of the highest-energy CRs should be linked with 
the powerful neutrino emission at a level at least comparable to the 
power in CRs. 

Despite a certain similarity to the standard (diffusive) acceleration 
mechanism, the converter mechanism violates some of its inherent 
relations. For example, the maximum particle energy attainable in the 
converter mechanism is $\Gamma^{1/2}$ times larger than in the 
standard one, provided acceleration occurs in a shear flow with 
chaotic ambient magnetic field. Also, the maximum energy of 
synchrotron photons appears to be $\Gamma^2$ times larger. 

Generally speaking, a beam pattern wider than $1/\Gamma$ is 
characteristic for any type of accompanying emission (synchrotron 
radiation at highest energies, neutrino emission or photons from pion 
decay), as well as for the escaping neutrons. This distinctive 
feature of the converter mechanism opens an interesting possibility 
for observation of the off-axis blazars and GRBs.

\begin{acknowledgments}
We thank John G. Kirk and Michal Ostrowski for useful discussions
and the referee for his suggestions.
This work was partially supported by the grants 02-02-16236 of the 
Russian Foundation for Basic Research and 00-15-96674 of the Council 
for Support of the Leading Scientific Schools, and by programs 
"Astronomia" of the Presidium of the Russian Academy of Science and 
of the Ministry of Industry, Science and Technology. 
\end{acknowledgments}

\end{document}